\documentclass{article}

\usepackage{graphics,graphicx}
\usepackage{pstricks,pst-node,pst-tree,pstricks-add}

\begin{document}

\title{A Functional Approach to \\ 
Standard Binary Heaps}
\author{VLADIMIR KOSTYUKOV \\
  \texttt{vladimir.v.kostyukov@gmail.com}}
\date{Dec 2013}
\maketitle

\begin{abstract}
This paper describes a new and purely functional implementation technique of binary heaps. A binary heap is a tree-based data structure that implements priority queue operations (insert, remove, minimum/maximum) and guarantees at worst logarithmic running time for them. Approaches and ideas described in this paper present a simple and asymptotically optimal implementation of immutable binary heap.
\end{abstract}

\section{Introduction}

There are several purely functional implementations of heaps such as \emph{Leftist Heap}, \emph{Binomial Heap} (Okasaki, 1999) and \emph{Braun Tree} (Braun and Rem, 1983), which are definitely the best choice as \emph{priority queues} in a functional setting. However, there are other heaps around without proper functional implementations. The simplest of them are standard \emph{binary heaps}, which do not fit well into a functional environment since their reference implementation is based on mutable arrays. This paper presents a new and purely functional implementation technique of binary heaps, with the same asymptotic bounds as in an imperative setting.

\section{Binary Heaps}

A \emph{binary heap} (Williams, 1964) is a data structure that implements priority queue interface and guarantees logarithmic running time for \verb"insert"/\verb"delete" operations and constant time access to \verb"minimum"/\verb"maximum" element. Binary heaps are commonly viewed as binary trees which satisfy two invariants:

\begin{enumerate}
    \item The \emph{shape} invariant: the tree is a complete binary tree.
    \item The \emph{min-heap} invariant: each node is less than or equal to each of its children.
\end{enumerate}

\section{Binary Heap Operations}

In Scala (Odersky et al., 2004) a binary min heap that holds an integer values might be represented as abstract \verb"Heap" class with two variants: \verb"Branch" and \verb"Leaf".
\begin{verbatim}

  abstract sealed class Heap {
    def min: Int
    def left: Heap
    def right: Heap
    def size: Int
    def height: Int
  }

  case class Branch(min: Int, left: Heap, right: Heap, 
                    size: Int, height: Int) extends Heap

  case object Leaf extends Heap {
    def min: Int = fail("Leaf.min")
    def left: Heap = fail("Leaf.min")
    def right: Heap = fail("Leaf.right")
    def size: Int = 0
    def height: Int = 0
  }

\end{verbatim}
Thus, using pattern matching with case classes, which are actually projections of \emph{Algebraic Data Types}, \verb"isEmpty" method can be written as

\begin{verbatim}

  def isEmpty: Boolean = this match {
    case Leaf => true
    case _ => false
  }

\end{verbatim}

Except for \verb"height" and \verb"size" operations, this signature looks like a functional implementation of \emph{binary search tree} (Okasaki, 1999). The two new operations are actually accessors to new fields in a heap - its height and size. This additional data should be accessible in constant time to define an efficient and simple \emph{search criteria} for \verb"insert" and \verb"remove" operations.

\section{Insertion in O(log n)}

Insertion into functional binary heap must not violate either of its invariants - neither the shape invariant nor the min-heap invariant. For this purpose two problems should be solved. First, to maintain the shape invariant a new node should be inserted in the first empty spot at the last level of the heap. Second, to maintain the min-heap invariant the inserted node should be \emph{bubbled up} to the heap root until it becomes greater than its parent.

Bubbling up is quite a simple transformation that can be done at each level in constant time. There are two cases depending on whether the violation is at left or right child. In both cases, the violation should be fixed by \emph{swapping} two nodes - the root node and the child that violates the min-heap invariant. There is also a third case, when it doesn't violate anything. In this case, a heap should be simply rebuilt with given parameters. In other words, all affected nodes should be copied in order to maintain data structure persistence. More precisely, \verb"bubbleUp" and \verb"insert" operations might be defined as

\begin{verbatim}

  def bubbleUp(x: Int, l: Heap, r: Heap): Heap = (l, r) match {
    case (Branch(y, lt, rt, _, _), _) if (x > y) => 
      Heap(y, Heap(x, lt, rt), r)
    case (_, Branch(z, lt, rt, _, _)) if (x > z) => 
      Heap(z, l, Heap(x, lt, rt))
    case (_, _) => Heap(x, l, r)
  }

  def insert(x: Int): Heap =
    if (isEmpty) Heap(x) 
    else if (???) bubbleUp(min, left.insert(x), right)
    else bubbleUp(min, left, right.insert(x))

\end{verbatim} , where the \emph{smart constructor} \verb"Heap" that creates a new singleton heap is defined as

\begin{verbatim}

  def Heap(x: Int, l: Heap = Leaf, r: Heap = Leaf): Heap = 
    Branch(x, l, r, l.size + r.size + 1, math.max(l.height, r.height) + 1)

\end{verbatim}

Note that \emph{height} of a heap is defined as max height of its children plus one, while \emph{size} of a heap is defined as sum of its children sizes plus one; and both are calculated only once in a heap constructor. Also, to simplify calculations, suppose that singleton heap's height is $1$.

The last thing to discuss is how to find a proper spot for a new node. This is actually a cornerstone of functional binary heaps. The main idea is based on two definitions of \emph{perfect binary trees}: \emph{math} and \emph{recursive}. Math definition: a perfect binary tree contains \emph{$2^{h + 1} - 1$} nodes, where \emph{$h$} is the height of the tree. Recursive definition: a tree is perfect if its children are perfect trees of the same height. Combining these facts together, one can define \emph{search criteria} which allow to fill a heap level by level from left to right, thereby maintaining the shape invariant. In other words, new nodes should be inserted in a way to make the heap be a perfect tree. This can be simply achieved by following requirements of the recursive definition, using the math definition as an efficient test on tree perfectness. Thus, the search criteria for insertion contain four cases depending on whether the children are perfect trees or not and whether their heights are equal or not.

\begin{verbatim}

  def insert(x: Int): Heap =
    if (isEmpty) Heap(x)
    else if (left.size < math.pow(2, left.height) - 1) 
      bubbleUp(min, left.insert(x), right)
    else if (right.size < math.pow(2, right.height) - 1) 
      bubbleUp(min, left, right.insert(x))
    else if (right.height < left.height) 
      bubbleUp(min, left, right.insert(x))
    else bubbleUp(min, left.insert(x), right)

\end{verbatim}
\begin{figure}
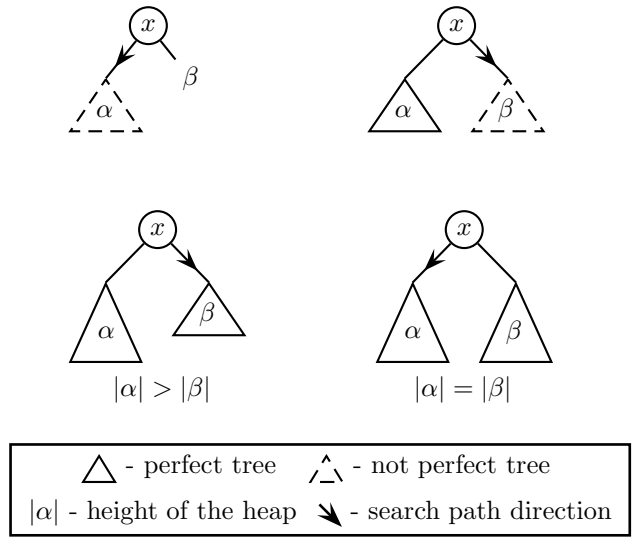

 \hspace{2.6cm}
 \rput[br](0.58,-1.22){$\alpha$}
 \pstree[levelsep=20pt,treesep=0.40]{\Tcircle{$x$}}
 {
  \psset{ArrowInside=->,arrowscale=2,ArrowInsidePos=0.7}
  \pstree{\Tp}{\Tfan[linestyle=dashed]}
  \psset{ArrowInside=none}
  \Tn
  \Tcircle[linestyle=none]{$\beta$}
 }
 \hspace{1.8cm}
 \rput[br](0.58,-1.22){$\alpha$} 
 \rput[br](1.94,-1.28){$\beta$} 
 \pstree[levelsep=20pt,treesep=0.40]{\Tcircle{$x$}}
 {
  \pstree{\Tp}{\Tfan}
  \psset{ArrowInside=->,arrowscale=2,ArrowInsidePos=0.7}
  \pstree{\Tp}{\Tfan[linestyle=dashed]}
  \psset{ArrowInside=none}
 }

 \vspace{1.0cm}

 \hspace{2.6cm}
 \rput[br](0.6,-1.44){$\alpha$} 
 \rput[br](1.94,-1.28){$\beta$} 
 \pstree[levelsep=20pt,treesep=0.40]{\Tcircle{$x$}}
 {
  \pstree[levelsep=30pt]{\Tp}{\Tfan}
  \psset{ArrowInside=->,arrowscale=2,ArrowInsidePos=0.7}
  \pstree{\Tp}{\Tfan}
  \psset{ArrowInside=none}
 }
 \hspace{1.5cm}
 \rput[br](0.6,-1.44){$\alpha$}
 \rput[br](1.94,-1.52){$\beta$} 
 \pstree[levelsep=20pt,treesep=0.40]{\Tcircle{$x$}}
 {
  \psset{ArrowInside=->,arrowscale=2,ArrowInsidePos=0.7}
  \pstree[levelsep=30pt]{\Tp}{\Tfan}
  \psset{ArrowInside=none}
  \pstree[levelsep=30pt]{\Tp}{\Tfan}
 }
 
 \rput[br](4.6,-0.5){$|\alpha| > |\beta|$}
 \rput[br](8.6,-0.5){$|\alpha| = |\beta|$}

 \vspace{1cm}

 \hspace{1.8cm}
 \psframebox[linewidth=1pt]{
  \begin{psmatrix}[rowsep=0.2cm]
   \trinode{A}{} - perfect tree \hspace{0.2cm} \trinode[linestyle=dashed]{A}{} - not perfect tree \hspace{0.2cm}\\
   $|\alpha|$ - height of the heap \hspace{0.2cm} 
   \psline[linewidth=1pt,arrowscale=2]{->}(0.3,0.2)(0.6,-0.1)  \hspace{0.6cm} - search path direction
   \end{psmatrix}
 }
 \vspace{0.4cm}
 \caption{Searching for the first empty spot in a heap.}
\end{figure}
The time complexity of \verb"insert" operation is $O(log\;n)$ since it requires to perform bubble up transformations for each node in a search path, and the longest possible path for complete trees is $log\;n$.

\section{Construction in O(n)}

Constructing binary heap from unordered input can be done in linear time (Floyd, 1964). Such performance is achieved by algorithm that constructs a complete heap in a bottom-up manner together with fixing all violations of the min-heap invariant. There is only one dangerous case that violates the min-heap invariant - the root node of new heap is greater than its children. This violation can be fixed by \emph{bubbling} wrongly placed node \emph{down} to the heap. The \verb"bubbleDown" operation can be written with pattern matching as
\begin{verbatim}

  def bubbleDown(x: Int, l: Heap, r: Heap): Heap = (l, r) match {
    case (Branch(y, _, _, _, _), Branch(z, lt, rt, _, _)) 
      if (z < y && x > z) => Heap(z, l, bubbleDown(x, lt, rt))
    case (Branch(y, lt, rt, _, _), _) 
      if (x > y) => Heap(y, bubbleDown(x, lt, rt), r)
    case (_, _) => Heap(x, l, r)
  }

\end{verbatim}

\begin{figure}
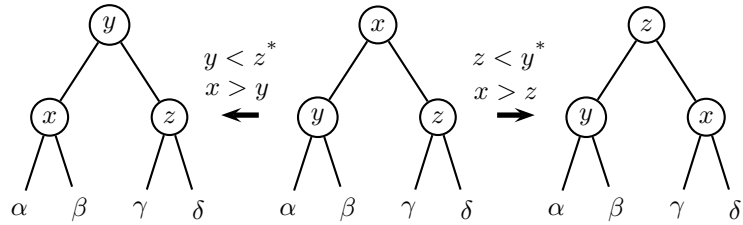

 \hspace{1cm}
 \pstree[levelsep=35pt,treesep=0.20]{\Tcircle{$y$}}
 {
  \pstree[levelsep=35pt]{\Tcircle{$x$}}{\Tcircle[linestyle=none]{$\alpha$} \Tcircle[linestyle=none]{$\beta$}} 
  \pstree[levelsep=35pt]{\Tcircle{$z$}}{\Tcircle[linestyle=none]{$\gamma$} \Tcircle[linestyle=none]{$\delta$}} 
 }
 \rput[br](0.66,-0.6){$y < z$\textsuperscript{*}} 
 \rput[br](0.54,-1){$x > y$} 
 \psline[linewidth=2pt]{<-}(-0.1,-1.2)(0.4,-1.2)
 \hspace{0.4cm}
 \pstree[levelsep=35pt,treesep=0.20]{\Tcircle{$x$}}
 {
  \pstree[levelsep=35pt]{\Tcircle{$y$}}{\Tcircle[linestyle=none]{$\alpha$} \Tcircle[linestyle=none]{$\beta$}} 
  \pstree[levelsep=35pt]{\Tcircle{$z$}}{\Tcircle[linestyle=none]{$\gamma$} \Tcircle[linestyle=none]{$\delta$}} 
 }
 \rput[br](0.66,-0.6){$z < y$\textsuperscript{*}} 
 \rput[br](0.54,-1){$x > z$} 
 \psline[linewidth=2pt]{->}(0,-1.2)(0.5,-1.2)
 \hspace{0.4cm}
 \pstree[levelsep=35pt,treesep=0.20]{\Tcircle{$z$}}
 {
  \pstree[levelsep=35pt]{\Tcircle{$y$}}{\Tcircle[linestyle=none]{$\alpha$} \Tcircle[linestyle=none]{$\beta$}} 
  \pstree[levelsep=35pt]{\Tcircle{$x$}}{\Tcircle[linestyle=none]{$\gamma$} \Tcircle[linestyle=none]{$\delta$}} 
 }
 
 \rput[br](9.0,-0.5){\textsuperscript{*} additional condition for bubbling down} 
 \vspace{0.5cm}
 \caption{Eliminating \emph{min-heap} invariant violations.}
\end{figure}

The \verb"heapify" operation constructs a \emph{complete} binary heap in a recursive way by using the ideas of array-based heaps representations - given the index $i$ of a node, the indices of its children - $2i + 1$ for left child and $2i + 2$ for right child. With inner function it looks like
\begin{verbatim}

  def heapify(a: Array[Int]): Heap = {
    def loop(i: Int): Heap = 
      if (i < a.length) bubbleDown(a(i), loop(2 * i + 1), loop(2 * i + 2))
      else Leaf

    loop(0)
  }

\end{verbatim}

The analysis of these operations is quite tricky. It seems that \verb"heapify"'s running time is $O(n\;log\;n)$, since each call to \verb"bubbleDown" costs $O(log\;n)$ and there are $O(n)$ such calls. This is correct for the upper bound, but it is not asymptotically tight. The thing is that running time of b\verb"bubbleDown" depends on the height of sub-heap it is applied to, and the heights of most sub-heaps are small. Thus, in most cases \verb"bubbleDown" runs in constant time and the total running time of heapify operation is $O(n)$. More detailed analysis can be found at (Cormen et al., 2001).

\section{Removal in O(log n)}

Removal is always pain in the neck for most of data structures. However, it is not so bad for binary heaps. There is quite a nice standard algorithm that allows to remove minimum/maximum from the heap in $O(log\;n)$ time. The algorithm joins two phases which maintain both invariants - replacing root node with last inserted one and bubbling it down. Considering that \verb"mergeChildren" is the first phase and \verb"bubbleRootDown" is the second one, \verb"remove" operation might be written as 
\begin{verbatim}

  def remove: Heap =
    if (isEmpty) fail("Empty heap.")
    else bubbleRootDown(mergeChildren(left, right))

\end{verbatim}
, where \verb"bubbleRootDown" can be defined as wrapper around \verb"bubbleDown" operation that was discussed previously.
\begin{verbatim}

  def bubbleRootDown(h: Heap): Heap = 
    if (h.isEmpty) Leaf
    else Heap.bubbleDown(h.min, h.left, h.right)

\end{verbatim}

The most interesting part of removal is its first phase. Quite a tricky problem should be solved at this phase: replacing the root of the heap with its last inserted node. First of all, it requires finding such node. This is what has already been done in insertion and all one needs to do is change the the search criteria a bit. The search criteria for removal contain four cases as well as for insertion, but it has a different meaning for some of them. First, if both node's children are empty, then this node is the last inserted one. Second, if both node's children are perfect trees and the left child is higher than the right one, then the last inserted node is somewhere on the left. Third, it is expected that the last inserted node is somewhere on the right if node's children are perfect trees of the same height.

When the last inserted node is found, it should be \emph{floated} to the place of the root. This can be done by using \emph{divide and conquer} algorithm with following ideas. Suppose that the last inserted node of the heap can be recursively floated to the place of its child's root. Then, depending on whether the child is left or right, its root should be lifted to the place of the heap's root. Finally, the affected child should be restored.

Combining all things together, \verb"mergeChildren" operation might be defined as 
\begin{verbatim}

  def mergeChildren(l: Heap, r: Heap): Heap = 
    if (l.isEmpty && r.isEmpty) Leaf
    else if (l.size < math.pow(2, l.height) - 1) 
      floatLeft(l.min, mergeChildren(l.left, l.right), r)
    else if (r.size < math.pow(2, r.height) - 1)
      floatRight(r.min, l, mergeChildren(r.left, r.right))
    else if (r.height < l.height)
      floatLeft(l.min, mergeChildren(l.left, l.right), r)
    else floatRight(r.min, l, mergeChildren(r.left, r.right))

  def floatLeft(x: Int, l: Heap, r: Heap): Heap = l match {
    case Branch(y, lt, rt, _, _) => Heap(y, Heap(x, lt, rt), r)
    case _ => Heap(x, l, r)
  }

  def floatRight(x: Int, l: Heap, r: Heap): Heap = r match {
    case Branch(y, lt, rt, _, _) => Heap(y, l, Heap(x, lt, rt))
    case _ => Heap(x, l, r)
  }

\end{verbatim}

The remove operation performs two walks along the search path of the heap. First, it searches for the last inserted node and floats it to the place of the root (maintaining the shape invariant). Second, it bubbles new root down (maintaining the min-heap invariant). Thus, keeping in mind that the longest possible path in a complete tree is $log\;n$, the total running time of \verb"remove" is $O(log\;n)$.

\section{Conclusion}

Functional setting brings some charm and beauty into data structures implementations. But it is not always possible to design a proper functional implementation that meets performance requirements. Sometimes, it is just close to impossible to convert a RAM-based algorithm into equivalent functional one - to make the mind think in terms of \emph{space} not \emph{time}. However, it can be done for binary heaps. The suggested implementation technique allows to achieve asymptotically optimal performance along with maintaining data structure persistence. And this is another good example in computer science that combines both elegant abstraction and witty implementation.

\bibliography{bibliography}

\begin{thebibliography}{}
 \bibitem[Okasaki, 1999]{Okasaki}
   Okasaki,~C. (1999), \emph{Purely functional data structures}, 
   Cambridge University Press.
 \bibitem[Braun and Rem, 1983]{Braun and Rem}
   Braun,~W. and Rem,~M. (1983) A logarithmic implementation of flexible arrays. \emph{Memorandum MR83/4}. Eindhoven University of Technology.
 \bibitem[Williams, 1964]{Williams}
   Williams,~J.~W.~J. (1964), Algorithm 232 - Heapsort, 
   \emph{Communications of the ACM} 7 (6): 347тАУ-348.
 \bibitem[Odersky et al., 2004]{Odersky}
   Odersky~M., Altherr~P., Cremet~V., Emir~B., Maneth~S., Micheloud~S., Mihaylov~N.,  Schinz~M., Stenman~E. and Zenger~M. (2004), An Overview of the Scala Programming Language, \emph{EPFL Technical Report IC/2004/64}, EPFL Lausanne.
 \bibitem[Floyd, 1964]{Floyd}
   Floyd,~Robert~W., (1964), Algorithm 245 - Treesort 3, 
   \emph{Communications of the ACM} 7 (12): 701
 \bibitem[Cormen et al., 2001]{Cormen}
   Cormen, ~T.~H., Leiserson,~C.~E. and Rivest,~R.~L. (2001), 
   \emph{Introduction to Algorithms} (2nd ed.), Cambridge, 
   Massachusetts: The MIT Press.
 \end{thebibliography}

\end{document}